\documentclass[%
 aip,
 amsmath,amssymb,
 reprint,%
]{revtex4-1}

\usepackage{graphicx}
\usepackage{dcolumn}
\usepackage{bm}

\usepackage[utf8]{inputenc}
\usepackage[T1]{fontenc}
\usepackage{mathptmx}
\usepackage{etoolbox}

\makeatletter
\def\@email#1#2{%
 \endgroup
 \patchcmd{\titleblock@produce}
  {\frontmatter@RRAPformat}
  {\frontmatter@RRAPformat{\produce@RRAP{*#1\href{mailto:#2}{#2}}}\frontmatter@RRAPformat}
  {}{}
}%
\makeatother
\begin{document}

\preprint{AIP/123-QED}

\title[Sample title]{Disorder enhanced exciton transport and quantum information spreading with the assistance of cavity QED}
\author{Weijun Wu}
    \affiliation{Department of Chemistry, Princeton University, Princeton, NJ 08540, U.S.}

\author{Ava N. Hejazi}
    \affiliation{Department of Chemistry, Princeton University, Princeton, NJ 08540, U.S.}

\author{Gregory D. Scholes}
    \email{gscholes@princeton.edu}
    \affiliation{Department of Chemistry, Princeton University, Princeton, NJ 08540, U.S.}

\date{\today}

\begin{abstract}
Molecular materials have been studied as a potential platform for highly efficient transport such as exciton transport and quantum information spreading. However, one detrimental factor to transport efficiency is the inherent disorder of the molecular system, where site-to-site hopping is suppressed by Anderson localization. Here we theoretically report a novel approach to eliminate the negative impact of disorder by strongly coupling the system to a cavity, where the cavity photon bridges spatially separated sites and builds an additional transport channel, cavity-mediated jumping. Our analysis of the open quantum system dynamics shows in terms of long-range transport, the two channels hold a competitive relation. When disorder suppresses site-to-site hopping, transport occurs mainly through cavity-mediated jumping in disguise. Therefore, with the assistance of the cavity, disorder in certain ranges can enhance transport and certain disordered systems can even be more efficient for transport than the homogeneous system. These results provide insight into the design of next-generation materials for exciton transport and quantum information spreading by leveraging hybrid light-matter states.
\end{abstract}

\maketitle

\section{\label{sec:level1} Introduction}
Energy transfer is a ubiquitous phenomenon in nature and critical to dynamics of the systems such as photocatalysts \cite{welin2017photosensitized, twilton2017merger, prier2013visible}, photosynthetic complexes \cite{emerson1932photochemical, mirkovic2017light}, single-molecular proteins \cite{roy2008practical, kobayashi2019bioluminescence}, and organic molecular aggregates devices \cite{hestand2018expanded, kasha1963energy, menke2014exciton, yoo2004efficient}, where quantum effects play a non-trivial role. A quantum mechanical description of the electronic energy transfer between molecules is exciton transport \cite{scholes2003long}, where the excited electronic states are treated as spatially bound electron-hole pairs (Frankel excitons \cite{spano2010spectral}) and the energy transfer is the quantum coherent process of exciton annihilation on one molecule and exciton creation on another molecule nearby
due to the dynamical polarization. 
Effective exciton transport occurs if the dipole-dipole interaction is stronger than the system-bath coupling.

Due to the quantum nature of exciton transport, it has been proposed that molecular systems favouring exciton transport can be potential platforms for quantum information processing \cite{wu2024foundations, wasielewski2020exploiting, sarovar2010quantum}, where each molecular site is considered as a qubit when only the excitation between the singlet states $|\mathrm{S}_0\rangle$ and $|\mathrm{S}_1\rangle$ is involved. Exciton transport is usually accompanied by quantum information spreading; efficient exciton transport implies high fidelity against the information spreading into the bath. Modifying the system locally at the molecular level can manipulate quantum information over long distances.

Thus, it is urgent to enhance the energy transfer efficiency of the materials, such as via morphology optimization \cite{tailor2020recent}. However, long range energy transfer may be inhibited by Anderson localization \cite{lagendijk2009fifty, evers2008anderson}, which results in the absence of diffusion in certain random lattices \cite{anderson1958absence}. Excitons scattered by the disordered potential behave as strong random-walk particles. Moreover, excitons are likely to decay during the random walk before any possible effective transport occurs. Disorder is a material intrinsic property, and thus Anderson localization is hard to eliminate. In this work, we focus in particular on the on-site energy disorder of a 1D molecular chain with nearest-neighbour hopping, where Anderson localization is a big challenge that suppresses single-particle transport.

\begin{figure}
    \includegraphics[width=0.48\textwidth]{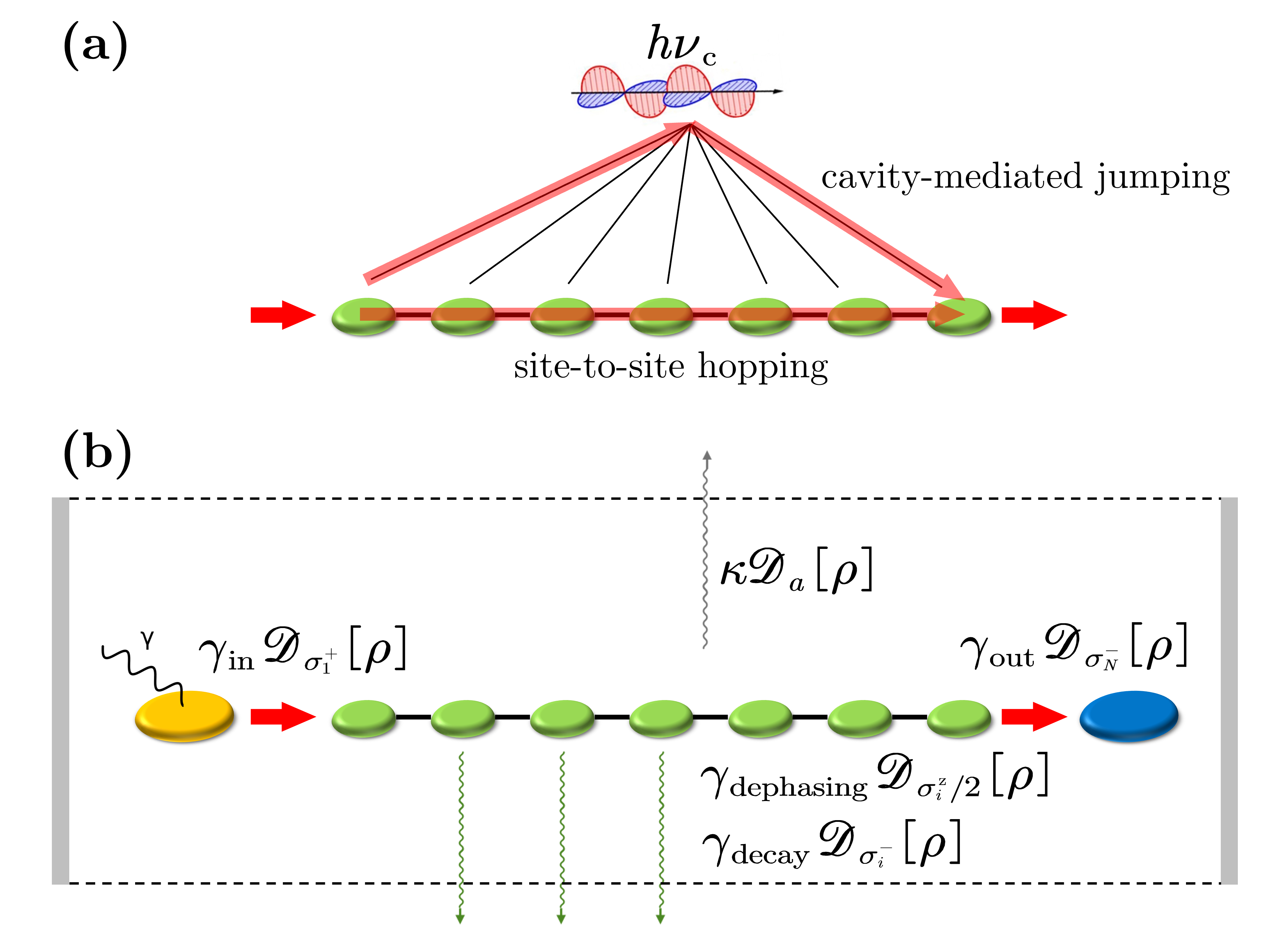}
    \caption{(a) Illustration for the site-to-site hopping channel and cavity-mediated jumping channel. (b) Illustration for the 1D molecular chain (green) with nearest-neighbour hopping inside the cavity, along with the external chromophore (yellow) and the reaction center (blue). Lindbladians are marked.}
    \label{fig:cavityQED}
\end{figure}

It has been proposed that cavity quantum electrodynamics (cavity QED) \cite{frisk2019ultrastrong,raimond2001manipulating,garcia2021manipulating, hagenmuller2017cavity, semenov2019electron, wu2022polariton} has the potential to enhance transport \cite{schachenmayer2015cavity, feist2015extraordinary, xu2023ultrafast, myers2018polariton}. When a layer of molecules is embedded into the Fabry–Pérot cavity (Fig.~\ref{fig:cavityQED}), strong light-matter interaction \cite{forn2019ultrastrong} and the corresponding quasiparticle, known as a polariton \cite{mandal2022theoretical, ebbesen2016hybrid}, emerge. Because the cavity mode coherently couples to many molecules, cavity QED induces collective coupling and multipartite entanglement \cite{wu2024molecular} that correlates the spatially distant molecules. Recent studies show that cavity-enhanced transport is robust to the presence of disorder \cite{suyabatmaz2023vibrational}, and disorder may not monotonically suppress transport \cite{allard2022disorder, chavez2021disorder, engelhardt2023polariton}. It is also interesting to see if disordered systems can be more beneficial for transport than a homogeneous system. Thus, it is important to have a deep understanding of how disorder affects the dynamical behaviour of the system coupled to the cavity, so as to increase the exciton transport and quantum information spreading without changing the materials' intrinsic properties.

In this work, we investigate the exciton transport and quantum information spreading for a 1D molecular chain with the Lindblad equation, by tuning the on-site energy disorder (characterised by standard deviation $w$) and collective light-matter coupling strength $G$. When the 1D chain is coupled to the cavity photon, the full system becomes quasi-1D. Apart from the site-to-site hopping in the bare 1D chain, the quasi-1D system induces another transfer channel, the cavity-mediated jumping (Fig.~\ref{fig:cavityQED}(a)). For a fixed $G$, the disorder can adjust the relative strength and competition relation between the two channels. Disorder in a certain range can suppress the site-to-site hopping and thus enhance the cavity-mediated jumping in disguise. Therefore, we demonstrate that in certain regions, increasing disorder can enhance transport, and furthermore, certain disordered systems can be more beneficial for exciton transport and quantum information spreading than the non-disordered system.

\section{\label{sec:level2} Model}
\subsection{\label{sec:level2.1} Cavity QED}
Here we consider the exciton-field interaction in the absence of spins.
The field is monochromatic of the fundamental frequency $\omega _{\mathrm{c}}=\mathrm{\pi} c/L$ regarding the cavity size $L$, and overtones are ignored because higher frequencies are far off-resonant with the transition frequency of the molecules: 
\begin{equation}
H_{\mathrm{C}}=\omega _{\mathrm{c}}\left( a^{\dagger}a+\frac{1}{2} \right)~. 
\label{eq:cavityHamiltonian}
\end{equation}
The exciton part, 
is considered as a 1D $N$-molecular aggregate 
with the quasiperiodic potential, 
described by the tight-binding model, due to the near-field dipole-dipole interaction:
\begin{equation}
H_{\mathrm{M}}=\sum_i^N{\omega _{\mathrm{i}}\frac{\sigma _{i}^{\mathrm{z}}}{2}}+J\sum_i^N{\left( \sigma _{i}^{+}\sigma _{i+1}^{-}+\sigma _{i}^{-}\sigma _{i+1}^{+} \right)}~.
\label{eq:moleculeHamiltonian}
\end{equation}
Each molecule is represented by Pauli matrices with the Frankel exciton energy disorder obeys the normal distribution: $\omega _i\sim \mathcal{N}\left( \omega _{\mathrm{m}},w^2 \right) $. The static disorder originates from the dynamic process of vibronic interaction with a long relaxation time. $J$ is the effective coupling strength between two nearest-neighbour molecules separated by $R$. Under the long-wavelength approximation 
($NR\ll\omega_{\mathrm{c}}^{-1}$)
and rotating wave approximation 
($\left| \omega _{\mathrm{c}}-\omega _{\mathrm{m}} \right| \sim G \ll \omega _{\mathrm{c}}+\omega _{\mathrm{m}}$), which are typical for optical cavities, the dipole-field interaction becomes
\begin{equation}
H_{\mathrm{int}}^{\mathrm{RWA}}=\frac{G}{\sqrt{N}}\sum_{i=1}^N{\left( a^{\dagger}\sigma _{i}^{-}+a\sigma _{i}^{+} \right)}~.
\label{eq:RWAHamiltonian}
\end{equation}
Note that dipole-dipole interaction and dipole field originate from the same general QED model but with different approximations, where the distinction between the two types of interaction can be demonstrated by the disorder.

In this paper, we study the Tavis-Cummins Hamiltonian \cite{tavis1968exact}, $H^{\mathrm{TC}}=H_{\mathrm{C}}+H_{\mathrm{M}}+H_{\mathrm{int}}^{\mathrm{RWA}}$
that effectively describs the electronic energy transport and quantum information spreading inside the cavity. The cavity and exciton are on resonance ($\omega _{\mathrm{m}}=\omega _{\mathrm{c}}=2.11\mathrm{eV}$). The Tavis-Cummins model has the U(1) symmetry and thus conserves the total excitation number $n=a^{\dagger}a+\frac{1}{2}\sum_i^N{\sigma _{i}^{\mathrm{z}}}$.


\subsection{\label{sec:level2.2} Open Quantum System Dynamics}
The dissipative exciton dynamics is studied by 
\begin{equation}
\frac{d\rho}{dt}=-\frac{i}{\hbar}\left[ H,\rho \right] +\sum_X{\gamma _X\mathcal{D} _X\left[ \rho \right]} \equiv \mathcal{L} \left[ \rho \right] ~,
\label{eq:Lindbald}
\end{equation}
where the non-unitary evolution, i.e. the Lindbladians, are $\gamma _X\mathcal{D} _X\left[ \rho \right] =\gamma _X\left( X\rho X^{\dagger}-\frac{1}{2}\left\{ X^{\dagger}X,\rho \right\} \right) $. Each single loss channel is weak enough to ensure the Markov dynamics, including the cavity loss ($X=a$, $\gamma _X=\kappa$), non-radiative electron-hole recombination of Frenkel exciton ($X=\sigma _{i}^{-}$, $\gamma _X=\gamma_\mathrm{decay}$) and exciton pure dephasing ($X=\sigma _{i}^{\mathrm{z}}/2$, $\gamma _X=\gamma_\mathrm{dephasing}$). Both photon and exciton thermal baths are assumed to be in $T\rightarrow0^{+}$ equilibrium. Collective dissipation such as superradiance \cite{huang2022room} is excluded from the Lindbladians because it is included in the Hamiltonian in a hermitian and non-Markov manner.

The exciton transport in the 1D chain is driven by an external source and drain (Fig.~\ref{fig:cavityQED}(b)). The first site of the aggregate is off-resonantly coupled to an external chromophore, which is continuously pumped by the laser. The last site is coupled to an external reaction center, where the energy is consumed by charge transfer and redox reactions. We do not discuss the details of the chemical dynamics within the chromophore and reaction center, but phenomenologically add two Lindbladians for the energy gain at the first site ($X=\sigma_{1} ^+$, $\gamma _X=\gamma_{\mathrm{in}}$) and energy loss at the last site ($X=\sigma_{N} ^-$, $\gamma_X=\gamma_{\mathrm{out}}$). The gain and loss rates are fixed to be low: $\gamma_{\mathrm{in}}=\gamma_{\mathrm{out}}=1\mathrm{THz}$. Because the source and drain are out-of-equilibrium, the Lindbladians effectively have the physical meaning: the chromophore at $T\rightarrow0^{-}$ and the reaction center at $T\rightarrow0^{+}$. So the exciton transport and information spreading is directional from the source to the drain in the absence of the inverse flow. 

Numerically, we use exact diagonalization (ED) to calculate the dynamics
\begin{equation}
\rho \left( t \right) =e^{\mathcal{L} t}\rho \left( 0 \right)~.
\label{eq:EDdynamics}
\end{equation}
The initial state $\rho \left( 0 \right)$ is the vacuum without any photon and exciton excitation, $|\mathrm{g}\rangle =|0\rangle \otimes |\mathrm{S}_0\rangle ^{\otimes N}$. Due to the low gain rate, the system decays from single-excited states to vacuum rapidly, preventing the population of mulit-excited states. So we can study the dynamics in the Hilbert space $\mathcal{H} ^{n \leqslant 1}$ spanned by $\left\{ |\mathrm{g}\rangle ,a^{\dagger}|\mathrm{g}\rangle ,\sigma _{1}^{+}|\mathrm{g}\rangle ,...,\sigma _{N}^{+}|\mathrm{g}\rangle \right\} $, with the dimension linearly dependent on $N$. This guarantees the feasibility of ED. The long-time limit, the unique steady state is built due to the balance of continuous pumping and dissipation, which can be calculated by
\begin{equation}
\mathcal{L} \left[ \rho \left( t\rightarrow \infty \right) \right] =0 ~.
\label{eq:EDsteady}
\end{equation}

\subsection{\label{sec:level2.3} Measurements and Observables}
\subsubsection{\label{sec:level2.3.1}coherence and population}
When the system is constrained to $\mathcal{H} ^{n\leqslant 1}$, coherence can be naturally defined in the basis $\left\{ |\mathrm{g}\rangle ,a^{\dagger}|\mathrm{g}\rangle ,\sigma _{1}^{+}|\mathrm{g}\rangle ,...,\sigma _{N}^{+}|\mathrm{g}\rangle \right\} $: 
\begin{equation}
    C_{\alpha, \beta}=\langle \mathrm{g}|O_{\alpha}{\rho O_{\beta}}^{\dagger}|\mathrm{g}\rangle ~,~~~~ \alpha,\beta\in \left\{ \mathrm{g},\mathrm{p},1,...,N \right\} ~,
    \label{eq:coherence}
\end{equation}
where $O_{\mathrm{g}}=I$, $O_{\mathrm{p}}=a$, $O_i=\sigma _{i}^{-}$. In the following of the paper, we denote $\alpha, \beta$ as the label for the basis states, and  $\mathrm{A}, \mathrm{B}$ as the label for two particles $\mathrm{A},\mathrm{B} \in \left\{\mathrm{p},1,...,N \right\}$. Because the $\rho \left( 0 \right) $ is incoherent and Lindbladians cannot generate the coherence, $C_{\mathrm{gA}}$ remains 0 during the evolution. Thus, coherence between two single-excited states $O_{\mathrm{A}}^{\dagger}|\mathrm{g}\rangle$ and $O_{\mathrm{B}}^{\dagger}|\mathrm{g}\rangle$ is base independent, and thus can be renamed as coherence between two particles A and B. In a more general sense, $C_{\mathrm{A,A}}$ is the population on particle A, and $C_{\mathrm{g,g}}$ is the vacuum population. 

\subsubsection{\label{sec:level2.3.2}exciton transport and energy current}
The gain and loss rate are low enough so that the linear response is valid and the system is assumed to be at a quasi-steady state all the time up to some linear correction. The total energy change rate is calculated by $d\mathrm{Tr}\left( H^{\mathrm{TC}}\rho \right) /dt=\sum_X{\gamma _X\mathrm{Tr}\left( H^{\mathrm{TC}}\mathcal{D} _X\left[ \rho \right] \right)}$, which implies that each Lindbladian linearly corresponds to an energy current\cite{wu2022polariton}. So the loss and gain currents can be defined as
\begin{align}
I_{\mathrm{out}}
&=
\gamma _{\mathrm{out}}\mathrm{Tr}\left( H^{\mathrm{TC}}\mathcal{D} _{\sigma _{N}^{-}}\left[ \rho \right] \right) ~,
\label{eq:currentout}
\\
I_{\mathrm{in}}
&=
\gamma _{\mathrm{in}}\mathrm{Tr}\left( H^{\mathrm{TC}}\mathcal{D} _{\sigma _{1}^{+}}\left[ \rho \right] \right) ~.
\label{eq:currentin}
\end{align}
Because of linear response, the energy current ratio, $I_{\mathrm{out}}/I_{\mathrm{in}}$, becomes a key measure of the efficiency of energy transport. When the system is constrained to $\mathcal{H} ^{n\leqslant 1}$, the current ratio has the explicit expression
\begin{equation}
    \frac{I_{\mathrm{out}}}{I_{\mathrm{in}}}=-\frac{\gamma _{\mathrm{out}}}{\gamma _{\mathrm{in}}}\frac{\omega _NC_{NN}+J\mathrm{Re}C_{N,N-1}+\frac{G}{\sqrt{N}}\mathrm{Re}C_{N,\mathrm{p}}}{\omega _1C_{\mathrm{gg}}} ~,
    \label{eq:currentratioexplicit}
\end{equation}
where coherence explicitly contributes to $I_{\mathrm{out}}$. 

Single exciton transport can also be described by diffusion, quantified by mean square displacement (MSD):
\begin{equation}
    \mathrm{MSD}=\sum_{i=1}^N{i^2C_{i,i}} ~,
    \label{eq:MSDExplicit}
\end{equation}
and the power law $\sqrt{\mathrm{MSD}}\propto t^{\nu}$ or $\nu=\frac{1}{2}\frac{\partial \ln \left( \mathrm{MSD} \right)}{\partial \ln t}$ indicates different types of diffusion models.

\subsubsection{\label{sec:level2.3.3}quantum information, entanglement and entropy}
Quantum information spreading can be measured based on the Rényi entropy \cite{renyi1961measures}. For example, the first-order Rényi entropy of a state is $S_1\left( \rho \right) =-\mathrm{Tr}\left( \rho \ln \rho \right) $. An important feature that distinguishes quantum information from classical information is quantum entanglement\cite{horodecki2009quantum, wu2024foundations}. In this paper, we focus on the pairwise entanglement between two particles, and the entanglement of the full system consisting of $N+1$ particles.

Bipartite entanglement is measured by the entanglement of formation \cite{wootters2001entanglement} $E_{\mathrm{F}}\left( \rho _{\mathrm{A,B}} \right) $. For a bipartite pure state, $E_{\mathrm{F}}$ is reduced to the entanglement entropy: $E_{\mathrm{F}}\left( |\psi \rangle _{\mathrm{A,B}} \right) =-\mathrm{Tr}\left( \rho _{\mathrm{A}}\ln \rho _{\mathrm{A}} \right)$, where $\rho _{\mathrm{A}}=\mathrm{Tr}_{\mathrm{B}}\left( \left( |\psi \rangle \langle \psi | \right) _{\mathrm{A,B}} \right) $. For a mixed bipartite state, $E_{\mathrm{F}}$ is the ensemble average of entanglement entropy, by minimizing over all the possible convex decomposition $\rho _{\mathrm{A,B}}=\sum_{i}{p_i\left( |\psi _i\rangle \langle \psi _i| \right) _{\mathrm{A,B}}}$:
\begin{equation}
E_{\mathrm{F}}\left( \rho _{\mathrm{A,B}} \right) =\underset{\left\{ p_i,|\psi _i\rangle _{\mathrm{A,B}} \right\}}{\min}\left\{ \sum_i{p_iE_{\mathrm{F}}\left( |\psi _i\rangle _{\mathrm{A,B}} \right)} \right\} 
\label{eq:EntanglementOfFormation}
\end{equation}
When the system is constrained to $\mathcal{H} ^{n\leqslant 1}$, $E_{\mathrm{F}}$ of a reduced density matrix of particles A and B has the closed form:
\begin{equation}
E_{\mathrm{F}}\left( \rho _{\mathrm{A},\mathrm{B}} \right) =h\left( \frac{1}{2}+\frac{1}{2}\sqrt{1-\left| 2C_{\mathrm{A},\mathrm{B}} \right|^2} \right)  ~,
\label{eq:EntanglementOfFormationExplicit}
\end{equation}
where $h\left( x \right) =-x\ln x-\left( 1-x \right) \ln \left( 1-x \right) $ is the Shannon entropy for the Bernoulli distribution. Eq.~(\ref{eq:EntanglementOfFormationExplicit}) show that entanglement of formation is explicitly related to coherence.

The entanglement measure of the full system is defined by the distance between the state $\rho$ and the set of separable states. When we take the distance measure to be the relative entropy, such measure obeys the entanglement monotone \cite{vedral1997quantifying} axiom:
\begin{equation}
E_{\mathrm{R}}\left( \rho \right) =\min_{\rho _{\mathrm{sep}}} \left\{ -\mathrm{Tr}\left( \rho \ln \frac{\rho _{\mathrm{sep}}}{\rho} \right) \right\} ~,
\label{eq:RelativeEntropyEntanglement}
\end{equation}
where the minimum is taken over all the separable states. When the system is constrained to $\mathcal{H} ^{n\leqslant 1}$, the separable states become incoherent and the relative entropy of entanglement has the closed form:\cite{sarovar2010quantum}
\begin{equation}
E_{\mathrm{R}}\left( \rho \right) =\mathrm{Tr}\left( \rho \ln \rho \right) -\sum_{\alpha\in \left\{ \mathrm{g},\mathrm{p},1,...,N \right\}}{C_{\alpha,\alpha}\ln C_{\alpha,\alpha}} ~.
\label{eq:RelativeEntropyEntanglementExplicit}
\end{equation}

\begin{figure*}
    \includegraphics[width=1\textwidth]{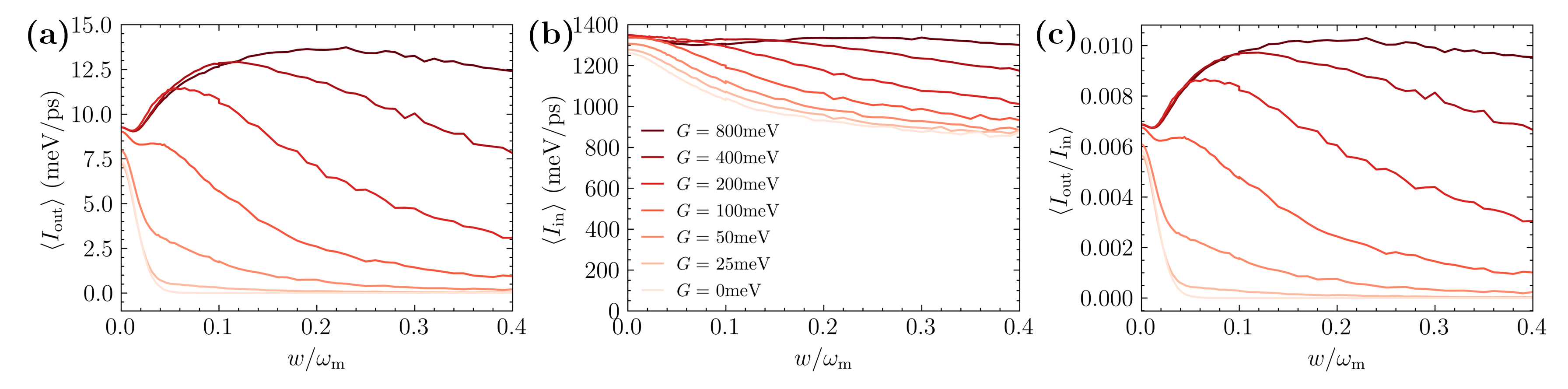}
    \caption{(a) Loss current $I_{\mathrm{out}}$, (b) gain current $I_{\mathrm{in}}$ and (c) current ratio $I_{\mathrm{out}}/I_{\mathrm{in}}$ of the steady state as a function of disorder $w/\omega _{\mathrm{m}}$ for the 1D chain with different collective coupling strength $G$ to the cavity, after ensemble average. Parameters: $N=40$, $\omega _{\mathrm{m}}=\omega _{\mathrm{c}}=2.11\mathrm{eV}$, $J=60\mathrm{meV}$, $\gamma _{\mathrm{decay}}=1.67\mathrm{THz}$, $\kappa=20\mathrm{THz}$, $\gamma _{\mathrm{dephasing}}=40\mathrm{THz}$, $\gamma _{\mathrm{in}}=\gamma _{\mathrm{out}}=1\mathrm{THz}$.}
    \label{fig:current}
\end{figure*}

\section{\label{sec:level3} Results and Discussion}
In this section, we discuss the numerical results of energy current and population dynamics, to derive the mechanism of disorder-enhanced transport, which originates from the competitive relationship between cavity-mediated jumping and site-to-site hopping. We further give more supporting examples including MSD and entropy calculation.

\subsection{\label{sec:level3.1}Energy Current}
Exciton dynamics usually occur on the ultrafast time scale. In the macroscopic time scale of a second, the system can be viewed as a steady state. FIG.~\ref{fig:current}(a)(b)(c) shows the steady state current as a function of $w$ and $G$, after ensemble averaging. In FIG.~\ref{fig:current}(c), when the system is outside the cavity ($G=0\mathrm{meV}$), the ratio drops to zero exponentially with the increase of $w$, demonstrating Anderson localization. When the system is inside the cavity, with the increase of G, the curves get higher, implying cavity-enhanced transport due to the establishment of the cavity-mediated jumping channel; A shoulder appears on finite $w$, which then becomes a peak in the ultrastrong coupling region. The existence of the peak indicates an optimal disorder $w_{\mathrm{opt}}\left(G\right)$ that maximizes the current ratio. When $w>w_{\mathrm{opt}}$, the system comes back to the conventional Anderson localization region where disorder suppresses transport, while the behaviour in the unconventional region ($w<w_{\mathrm{opt}}$) is interesting, where Anderson localization seems to fail and disorder dramatically enhances transport. With the increase of $G$, the peak gets higher, wider, and right-shifted (We further report $w_{\mathrm{opt}}\left( G \right) \propto G$). Thus, when the 1D chain is ultra-strongly coupled to the cavity, the systems with on-site energy disorders in a broad range are more efficient for energy transfer than the homogeneous system.

\begin{figure*}
    \includegraphics[width=1\textwidth]{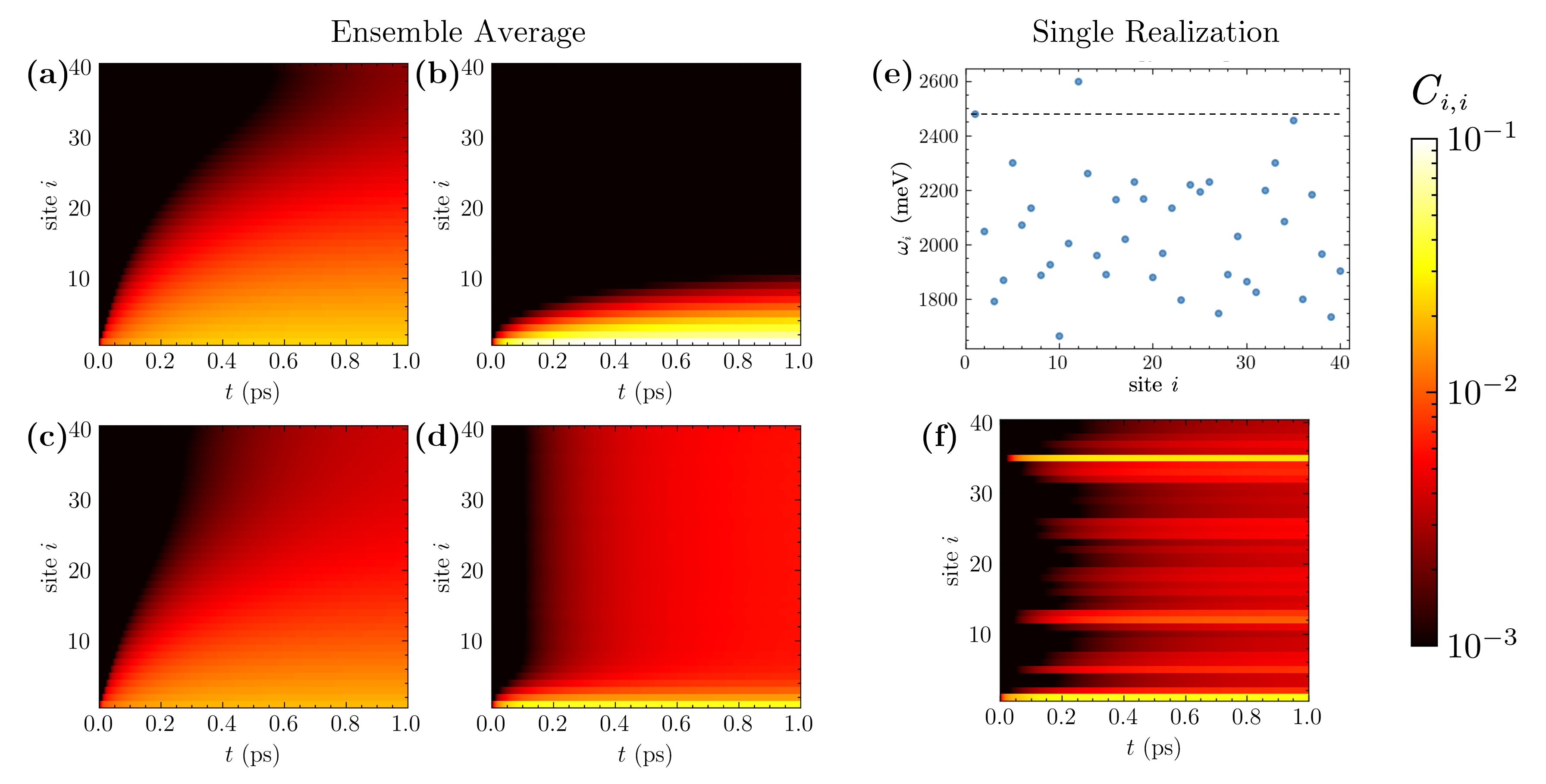}
    \caption{(a)(b)(c)(d) are the population dynamics $C_{i,i}\left( t \right) $ for each site $i$, after ensemble average. The four cases are (a) homogeneous and outside cavity ($w/\omega _{\mathrm{m}}=0, G=0\mathrm{meV}$), (b) disordered and outside cavity ($w/\omega _{\mathrm{m}}=0.1, G=0\mathrm{meV}$), (c) homogeneous and inside cavity ($w/\omega _{\mathrm{m}}=0, G=400\mathrm{meV}$), (d) disordered and inside cavity ($w/\omega _{\mathrm{m}}=0.1, G=0\mathrm{meV}$). For case D, a special realization of (e) the on-site energy $\omega_i$ is taken with (f) the corresponding population dynamics. All the population data are displayed in the heat map of the same scale. Parameters: same as FIG.~\ref{fig:current}.
    }
    \label{fig:populationdynamics}
\end{figure*}

\begin{figure}
    \includegraphics[width=0.48\textwidth]{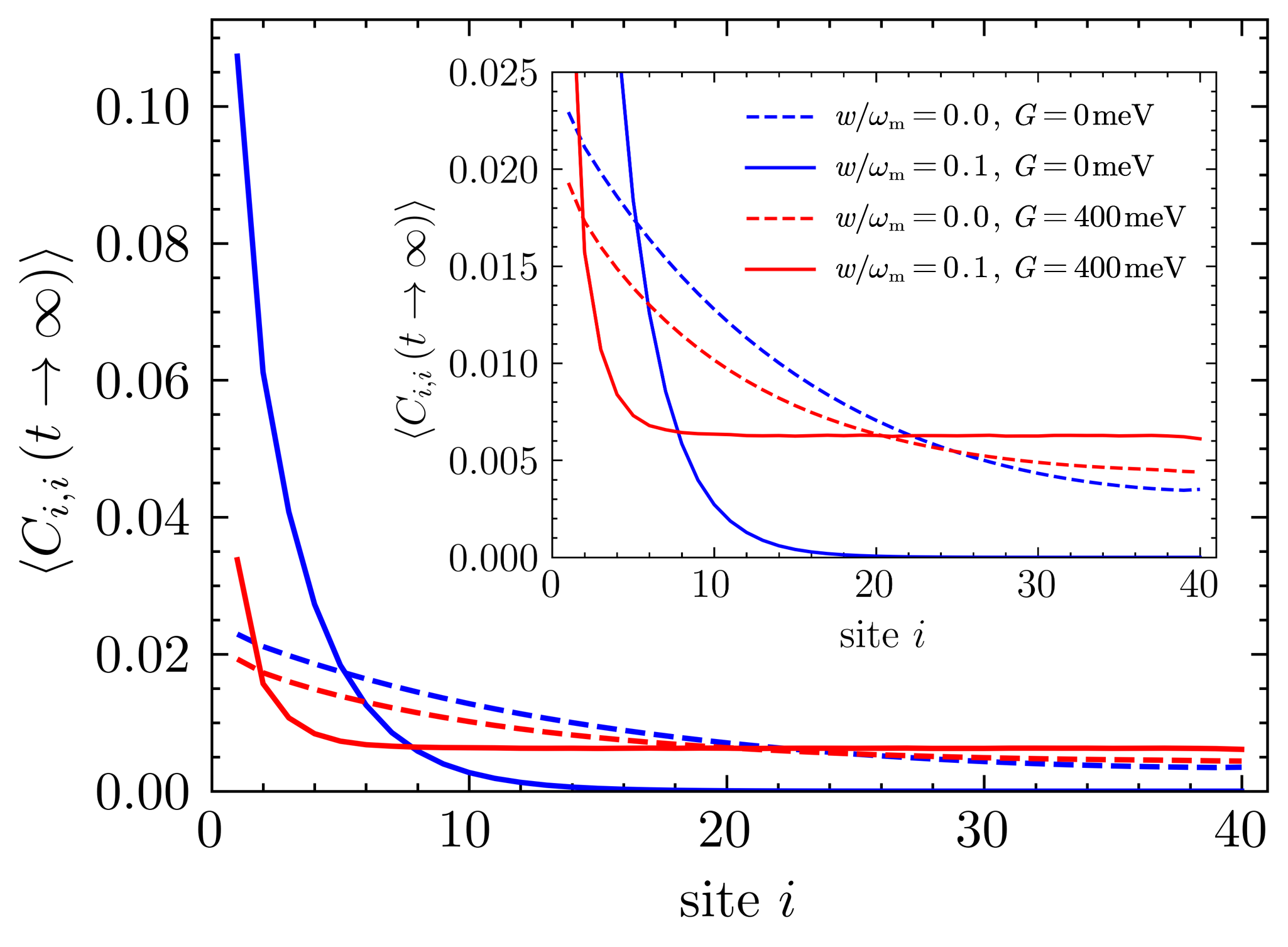}
    \caption{Steady state population distribution on each site. Inset: a zoomed-in figure for the low-population region. Parameters: same as FIG.~\ref{fig:current}. 
    }
    \label{fig:steadystatepopulation}
\end{figure}

Note that FIG.~\ref{fig:current}(a) has similar trends as FIG.~\ref{fig:current}(c), because the curves of $\left< I_{\mathrm{in}} \right> $ with respect to $w$ is relatively flat (FIG.~\ref{fig:current}(c)). Thus, the loss current flowing into the drain also measures the transport. As shown in Eq.~(\ref{eq:currentratioexplicit}), the current is mainly contributed by the population of site $N$. To reveal the current behaviour, we need to study the population dynamics. We take $G=400\mathrm{meV}$ as an example. The corresponding optimal disorder is $w_{\mathrm{opt}}/\omega_{\mathrm{m}}=0.1$, as shown in FIG.~\ref{fig:current}(c), which is a typical value for chromophore absorption broadening. Next, we focus specifically on four cases: (A) homogeneous and outside cavity ($w/\omega _{\mathrm{m}}=0, G=0\mathrm{meV}$), (B) disordered and outside cavity ($w/\omega _{\mathrm{m}}=0.1, G=0\mathrm{meV}$), (C) homogeneous and inside cavity ($w/\omega _{\mathrm{m}}=0, G=400\mathrm{meV}$), (D) disordered and inside cavity ($w/\omega _{\mathrm{m}}=0.1, G=400\mathrm{meV}$). 


\subsection{\label{sec:level3.2}Population Dynamics}
FIG.~\ref{fig:populationdynamics} shows the population dynamics of each molecular site $\left< C_{ii}\left( t \right) \right> $, after ensemble averaging. Homogeneous and bare 1D chain (case A, FIG.~\ref{fig:populationdynamics}(a)) exhibits typical exciton diffusion dynamics due to site-to-site hopping. When disorder is added to the bare 1D chain (case B, FIG.~\ref{fig:populationdynamics}(b)), Anderson localization dramatically suppresses site-to-site hopping and the population is trapped on the first few sites. When the homogeneous 1D chain is coupled to the cavity (case C, FIG.~\ref{fig:populationdynamics}(c)), exciton transport is similar to FIG.~\ref{fig:populationdynamics}(a) but more ballistic due to cavity-mediated jumping. Thus, the exciton can arrive at site $N$ in a shorter time compared to case A, demonstrating the cavity-enhanced transport. All cases A, B, and C show exciton transport stepwise, a feature of site-to-site hopping.

When the disordered 1D chain is coupled to the cavity (case D, FIG.~\ref{fig:populationdynamics}(d)), the population dynamics shows a very different pattern. At the early time ($t<0.1\mathrm{ps}$), the population is mainly localized at the first few sites, similar to FIG.~\ref{fig:populationdynamics}(b). At $t>0.1\mathrm{ps}$, the population is equally distributed over all the sites other than the first few, implying synchronization and simultaneous behaviour beyond step-wise transport. Thus, the site $N$ holds a significant exciton population in a very short time. To understand this non-trivial pattern of dynamics, we check a specific energy configuration (Fig.~\ref{fig:populationdynamics}(e)). In this single realization, by coincidence, site 1 is only resonant to site 35, and thus the exciton can effectively transport from site 1 to site 35 (Fig.~\ref{fig:populationdynamics}(f)) via cavity-mediated jumping, without passing through the off-resonant sites in between. Then the exciton only needs to transport from site 35 to site $N$, which is a short-range site-to-site hopping feasible under Anderson localization. 
Thus, in case D, the energy difference is more important than the distance in terms of effective exciton transport, and disorder won't suppress cavity-mediated jumping as long as resonance occurs. Back to the ensemble average where each site has equal probabilities of being resonant to site 1, the synchronization emerges as shown in FIG.~\ref{fig:populationdynamics}(d) and case D favours long-range exciton transport the most. 

We also discuss the late-time behaviour of the population distribution of each site after ensemble averaging, (FIG.~\ref{fig:steadystatepopulation}). Homogeneous and bare 1D chain (case A, blue dashed line) shows a typical gradient population distribution along the 1D chain. When disorder is added to the bare 1D chain (case B, blue solid line), exciton transport is short-ranged: the first 4 sites are much more populated than case A, while no population on sites beyond site 15. When the homogeneous 1D chain is coupled to the cavity (case C, red dashed line), the population distribution is similar to case A (blue dashed line), but with less population on sites 1-25 and more population on sites 25-$N$, because the cavity-mediated jumping favours long-range transport and transports more population to the far sites. 

When the disordered 1D chain is coupled to the cavity (case D, red solid line), the population distribution is similar to case B (blue solid line) but decays to a finite value on the far sites, which again demonstrates the cavity-enhanced transport due to the cavity-mediated jumping. The population on the site farther than site 10 is nearly constant because on average, all the sites have the same probability of being resonant to the site 1. Compared to the homogeneous cavity system (cases C, red dashed lines), case D has less population on sites 2-20 but more population on sites 24-$N$ (FIG.~\ref{fig:steadystatepopulation} inset). This is because the population on site 1 is more likely to go through the relatively strong cavity-mediated jumping channel when site-to-site hopping is weak, which lowers the near-site population but increases the far-site population. In contrast, in case C, site-to-site hopping coexists with cavity-mediated jumping and transfers the macroscopic population to the near sites.

\begin{figure*}
    \includegraphics[width=1\textwidth]{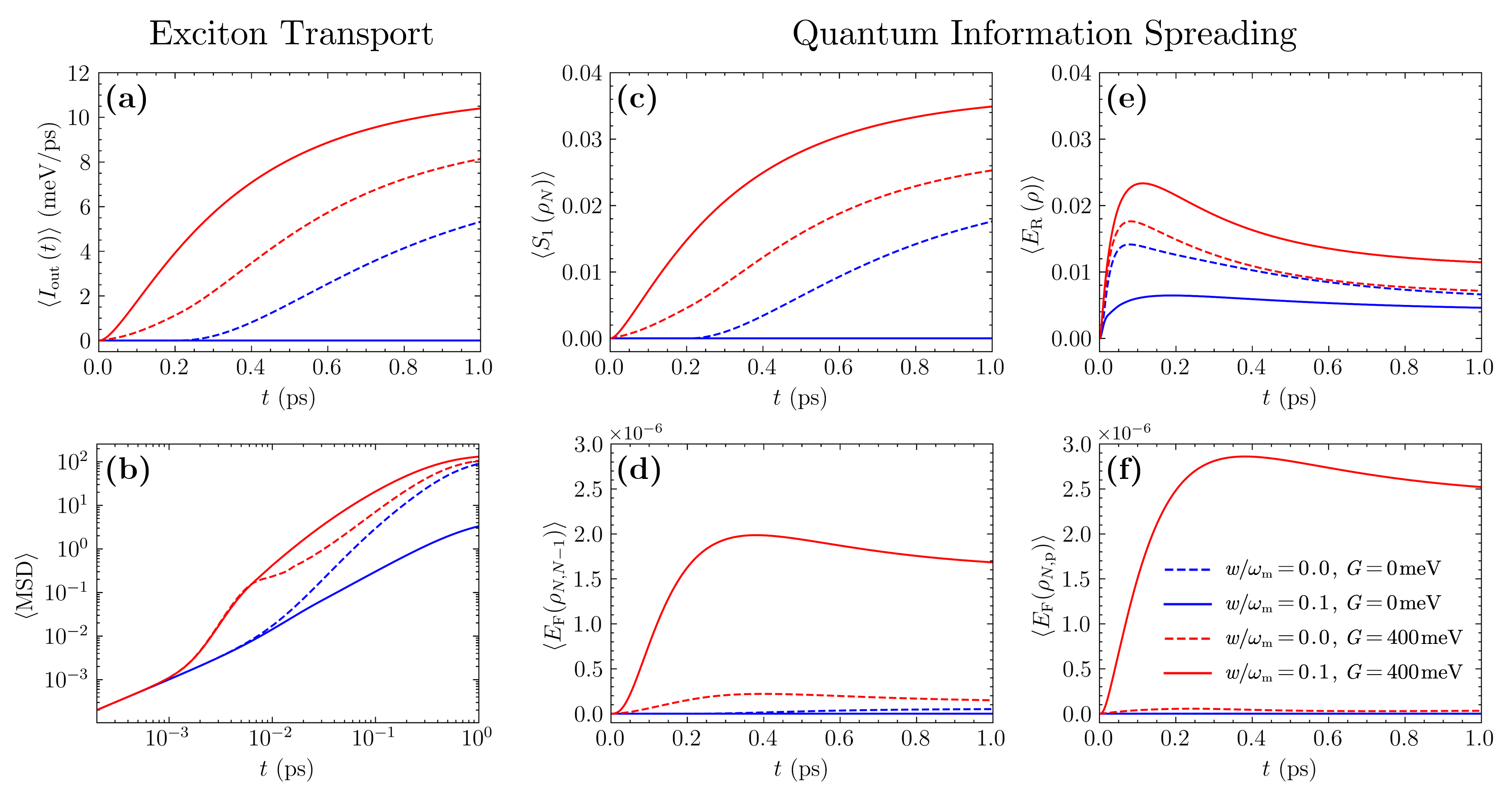}
    \caption{Exciton transport: time evolution of (a) loss current and (b) mean square displacement for the four cases, after ensemble average. Quantum information spreading: time evolution of (c) first-order Rényi entropy of site $N$, (d) entanglement of formation between site $N$ and site $N-1$, (e) relative entropy of entanglement of the full system, (f) entanglement of formation between site $N$ and cavity photon, of the four cases, after ensemble average. Parameters: same as FIG.~\ref{fig:current}. 
    }
    \label{fig:dynamics4cases}
\end{figure*}

\subsection{\label{sec:level3.3}Mechanism: Competition between Two Chennals}

Therefore, we interpret that the two transfer channels, site-to-site hopping and cavity-mediated jumping, hold a competitive relationship in terms of $\left<C_{N,N}\right>$. Site-to-site hopping favours near sites and short-range transport, while cavity-mediated jumping treats all the sites equally, without any preference (for single realization, there is a preference for those sites resonant to site 1, but after ensemble average, there is no preference). Therefore, cavity-mediated jumping is more beneficial for the far sites and long-range transport than site-to-site hopping. In the absence of disorder (case C), both transfer channels are strong, and we observe the gradient population distribution because site-to-site hopping prefers the near sites. When disorder is included (case D), site-to-site hopping is weak due to Anderson localization, but cavity-mediated jumping can still be strong because of the occasional resonance. Because cavity-mediated jumping is, therefore, the preferred channel in this case, the population is more likely to transfer to the far sites, resulting in disorder-enhanced transport (FIG.~\ref{fig:steadystatepopulation} inset, right ends). 

In contrast, when in terms of the population on the first site $\left<C_{1,1}\right>$, site-to-site hopping and cavity-mediated jumping hold a cooperative relationship. When disorder eliminates the site-to-site hopping, the population on site 1 can only be transferred through cavity-mediated jumping and thus more population remains on site 1. So disorder always suppresses the population leaving site 1, regardless of the coupling to the cavity (FIG.~\ref{fig:steadystatepopulation}, left ends). In fact, Our theory verifies Anderson localization rather than disobeying it.

\subsection{\label{sec:level3.4}Exciton Transport and Quantum Information Spreading}
By revealing the mechanism of disorder-enhanced transport with the assistance of cavity QED, we further study the other measures of exciton transport and quantum information spreading. In FIG.~\ref{fig:dynamics4cases}(a), case D (red solid line) maintains the largest $I_{\mathrm{out}}$ over all time, indicating that disorder-enhanced transport is not merely a late time effect, but also shows up in the early time ($t<0.1\mathrm{ps}$, see FIG.~\ref{fig:populationdynamics}). This implies that populating cavity photons is sufficiently rapid so that the cavity-mediated jumping channel can be built in a short time. In FIG.~\ref{fig:dynamics4cases}(b), case B (blue solid line) obeys the typical diffusion model that $\sqrt{\mathrm{MSD}}\propto t^{1/2}$, while the other three cases start with diffusion but later raise to super-diffusion. The super-diffusion drops to the sub-diffusion after a certain time because of the boundary effect, so MSD at an early time is a good measure for the bulk. Case D (red solid line) shows the largest MSD and the ballistic feature ($\nu=1$) for the longest time. In total, we conclude that the disordered quasi-1D chain is the most beneficial for exciton transport both on the edge and in the bulk.

In FIG.~\ref{fig:dynamics4cases}(c), the highest entropy of site $N$ shows up in case D (red solid line), indicating the information spreads most rapidly and site $N$ carries the most information in case D. Apart from the information spreading to the edge, we also focus on the global information and the multipartite entanglement, where higher entanglement implies a stronger correlation among the particles, which is beneficial for quantum information spreading in the bulk. In FIG.~\ref{fig:dynamics4cases}(d), the system in case D (red solid line) is the most entangled thoughout the dynamics, where the relative entropy of entanglement $E_{\mathrm{R}}\left( \rho \right)$ increases dramatically at early time ($t<0.1\mathrm{ps}$), so that the quantum information spreads the most coherently and effectively. After a certain time, $E_{\mathrm{R}}\left( \rho \right)$ drops and converges, as expected from boundary effects in a finite system. (similar to MSD in FIG.~\ref{fig:dynamics4cases}(b)), where dissipation lowers the entanglement. In total, we conclude that the disordered cavity system is the most beneficial for quantum information spreading both on the edge and in the bulk. 

Eq.~(\ref{eq:currentratioexplicit}) indicates the contribution of quantum coherence $C_{N,\mathrm{p}}$ and $C_{N,N-1}$ to the current, which corresponds to the two channels. Inspired by this, we check the bipartite entanglement, quantified by entanglement of formation, between site $N$ and cavity mode (FIG.~\ref{fig:dynamics4cases}(f)) and between site $N$ and site $N-1$ (FIG.~\ref{fig:dynamics4cases}(d)), to see their contribution to quantum information spreading. In case D (red solid line), both $E_{\mathrm{F}}\left( \rho _{N,\mathrm{p}} \right) $ and $E_{\mathrm{F}}\left( \rho _{N,N-1} \right) $ are much larger than the ones in the other three cases, meaning that quantum information spreading through both cavity-mediated jumping and site-to-site hopping to site $N$ are the most effective in case D. Note $E_{\mathrm{F}}\left( \rho _{N,\mathrm{p}} \right) >E_{\mathrm{F}}\left( \rho _{N,N-1} \right) $ in case D (red solid line), while this relation is inverse in case C (red dashed line). This also implies that in case D, cavity-mediated jumping plays a more important role than site-to-site hopping to enhance quantum information spreading.

\section{\label{sec:level4} Conclusion}
To summarize, we propose that energetic disorder can enhance transport within a 1D chain when the system is coupled to the cavity: in a certain range of disorders, increasing disorder can enhance long-range transport, and furthermore, certain disordered systems can be more efficient for transport than the non-disordered system. This seems to disobey the intuition that disorder is detrimental to transport due to Anderson localization. 

To study the quantum dynamics properties of transport for a 1D chain coupled to the cavity, i.e. a quasi-1D system, we build the open quantum system model to drive transport by continuously pumping the first site and extracting energy from the last site. 
The current ratio $\left< I_{\mathrm{out}}/I_{\mathrm{in}} \right> $ as a function of disorder $w$ for a fixed collective light-matter coupling $G$ shows a peak higher than the homogeneous $w=0$ point (FIG.~\ref{fig:current}(c)), indicating disorder-enhanced transport with the assistance of cavity QED. The optimal disorder $w_{\mathrm{opt}}$ seems to be proportional to $G$. We then proposed the mechanism of disorder-enhanced transport, due to the competition between the two possible transport channels: site-to-site hopping and cavity-mediated jumping. Site-to-site hopping favours near sites and short-range transport, while cavity-mediated jumping treats all the sites equally and is thus more beneficial for the far sites and long-range transport than site-to-site hopping. When disorder is added, site-to-site hopping is weak due to Anderson localization, but cavity-mediated jumping can be still strong as long as occasional resonance occurs (FIG.~\ref{fig:populationdynamics}(f)), so that population is likely to transfer to the far sites via the strong cavity-mediated jumping channel.

We then study some other features of exciton transport (current flowing into the drain, mean squared displacement) and quantum information spreading (first-order Rényi entropy, relative entropy of entanglement and entanglement of formation). All these quantities are optimized by the disordered quasi-1D chain. Thus, we claim disorder-enhanced exciton transport and quantum information spreading
with the assistance of cavity QED.

We emphasize that our prediction is important for the optimization of the materials that rely on the transport properties. Since disorder is ubiquitous, we not only eliminate the detrimental effect of disorder but also utilize disorder by coupling the materials to the cavity to enhance transport, without changing the material's intrinsic properties. Our proposal offers a novel way to design organic devices and new energy materials for energy conversion, as well as manipulate quantum information and entanglement with high fidelity against the environment over a long distance. It is also interesting to consider the realization of this mechanism in other quantum systems such as trapped ions \cite{monroe2021programmable} where phonons induce the bosonic-mediated long-range interaction.

Returning to the foundational significance of disorder-enhanced transport, our theory does not disobey Anderson localization but reveals a new pathway by which to exploit single-particle localization. Our system is quasi-1D, where an indirect all-to-all coupling is achieved by the cavity, so that Anderson localization facilitates the resonant transport pathway but selectively eliminates the off-resonant decay channel. When the disorder is too large, Anderson localization still suppresses the transport. Therefore, it is interesting to understand the disorder-enhanced transport in the many-body system, where many-body localization phase transition \cite{nandkishore2015many} may not occur and quantum thermalization may lead to transport in quasi-1D system in a different pattern.

\section*{Supplementary Material}
See Supplementary Material for theoretical details of entropy and more data on the dynamics.

\begin{acknowledgments}
This research is funded by the U.S. Department of Energy, Office of Science, Office of Basic Energy Sciences Solar Photochemistry program, under Award Number DE-SC0015429. The authors acknowledge Abraham Nitzan, David A. Huse, Biao Lian, David R. Reichman and Christopher Jarzynski for the valuable discussion and review.
\end{acknowledgments}

\section*{AUTHOR DECLARATIONS}
\subsection*{Conflict of Interest}
The authors have no conflicts to disclose.

\section*{Data Availability Statement}
The data that support the findings of this study are available
from the corresponding author upon reasonable request.

\nocite{*}
\bibliography{aipsamp}

\onecolumngrid
\newpage 

\setcounter{section}{0}
\setcounter{equation}{0}
\setcounter{figure}{0}
\renewcommand\thesection{S\arabic{section}}
\renewcommand \thefigure {S\arabic{figure}}

\begin{center}
{\Large Supplementary Material \\ 
\vspace{0.2cm}
Disorder enhanced exciton transport and quantum information spreading with the assistance of cavity QED
}
\end{center}

\section{Theoretical Details}
\subsection{First-order Rényi entropy}
The reduced density matrix of particle A can be calculated by tracing out the other particles
\begin{equation}
    \rho _{\mathrm{A}}=C_{\mathrm{A},\mathrm{A}}O_{\mathrm{A}}^{\dagger}|\mathrm{g}\rangle \langle \mathrm{g}|O_{\mathrm{A}}+C_{\mathrm{g},\mathrm{A}}|\mathrm{g}\rangle \langle 1|+C_{\mathrm{A},\mathrm{g}}|1\rangle \langle \mathrm{g}|+\left( 1-C_{\mathrm{A},\mathrm{A}} \right) |\mathrm{g}\rangle \langle \mathrm{g}|=\left( \begin{matrix}
	1-C_{\mathrm{A},\mathrm{A}}&		C_{\mathrm{g},\mathrm{A}}\\
	C_{\mathrm{A},\mathrm{g}}&		C_{\mathrm{A},\mathrm{A}}\\
\end{matrix} \right) 
~,
\label{eqS:ReducedDMA}
\end{equation}
By diagonalizing the matrix, we have 
\begin{equation}
    S_{\mathrm{A}}=-\rho _{\mathrm{A}}\ln \rho _{\mathrm{A}}=h\left( p_{\mathrm{A}} \right) ~,
\label{eqS:entropy}
\end{equation}
where $h\left( x \right) =-x\ln x-\left( 1-x \right) \ln \left( 1-x \right) $ and $p_{\mathrm{A}}=\frac{1}{2}-\sqrt{\left( \frac{1}{2}-C_{\mathrm{A},\mathrm{A}} \right) ^2+\left| C_{\mathrm{A},\mathrm{g}} \right|^2}$. Considering the fact that coherence is very small, we have $S_{\mathrm{A}}\approx h\left( C_{\mathrm{AA}} \right) $.

\subsection{Entanglement of formation}
We first calculate concurrence, which is an entanglement measure for bipartite qubits entanglement. Concurrence was first proposed by \citet{hill1997entanglement} based on the fidelity operator $R=\sqrt{\sqrt{\rho _{\mathrm{AB}}}\tilde{\rho}_{\mathrm{AB}}\sqrt{\rho _{\mathrm{AB}}}}$, where $\tilde{\rho}_{\mathrm{AB}}=\left( \sigma _{\mathrm{A}}^{\mathrm{y}}\sigma _{\mathrm{B}}^{\mathrm{y}} \right) \rho _{\mathrm{AB}}^{*}\left( \sigma _{\mathrm{A}}^{\mathrm{y}}\sigma _{\mathrm{B}}^{\mathrm{y}} \right) $ and $\rho _{\mathrm{AB}}^{*}$ is the complex conjugate of $\rho _{\mathrm{AB}}$ in $\sigma^{\mathrm{z}}$ basis. $R$ is Hermitian and has the four real eigenvalues $\lambda _{m}$ in descending order. Concurrence is defined as
\begin{equation}
E_{\mathrm{C}}\left( \rho _{\mathrm{AB}} \right) =\max \left\{ 0,\lambda _1-\lambda _2-\lambda _3-\lambda _4 \right\} 
\label{equ:concurrence}
\end{equation}

The reduced density matrix of particle A and B can be calculated by tracing out the other particles
\begin{equation}
    \rho _{\mathrm{A},\mathrm{B}}=\left( 1-C_{\mathrm{A},\mathrm{A}}-C_{\mathrm{B},\mathrm{B}} \right) |\mathrm{g}\rangle \langle \mathrm{g}|+C_{\mathrm{A},\mathrm{A}}O_{\mathrm{A}}^{\dagger}|\mathrm{g}\rangle \langle \mathrm{g}|O_{\mathrm{A}}+C_{\mathrm{A},\mathrm{B}}O_{\mathrm{A}}^{\dagger}|\mathrm{g}\rangle \langle \mathrm{g}|O_{\mathrm{B}}+C_{\mathrm{B},\mathrm{B}}O_{\mathrm{B}}^{\dagger}|\mathrm{g}\rangle \langle \mathrm{g}|O_{\mathrm{B}}+C_{\mathrm{B},\mathrm{A}}O_{\mathrm{B}}^{\dagger}|\mathrm{g}\rangle \langle \mathrm{g}|O_{\mathrm{A}}~.
\label{eqS:ReducedDMAB}
\end{equation}
So we have
\begin{equation}
    \rho _{\mathrm{A},\mathrm{B}}=
    \left( \begin{matrix}
    	1-C_{\mathrm{A},\mathrm{A}}-C_{\mathrm{B},\mathrm{B}}&		0&		0&		0\\
    	0&		C_{\mathrm{A},\mathrm{A}}&		C_{\mathrm{A},\mathrm{B}}&		0\\
    	0&		C_{\mathrm{B},\mathrm{A}}&		C_{\mathrm{B},\mathrm{B}}&		0\\
    	0&		0&		0&		0\\
    \end{matrix} \right) 
\end{equation}

\begin{equation}
\rho _{\mathrm{A},\mathrm{B}}^{*}=\left( \begin{matrix}
	1-C_{\mathrm{A},\mathrm{A}}-C_{\mathrm{B},\mathrm{B}}&		0&		0&		0\\
	0&		C_{\mathrm{A},\mathrm{A}}&		C_{\mathrm{B},\mathrm{A}}&		0\\
	0&		C_{\mathrm{A},\mathrm{B}}&		C_{\mathrm{B},\mathrm{B}}&		0\\
	0&		0&		0&		0\\
\end{matrix} \right) 
\end{equation}

\begin{equation}
\tilde{\rho}_{\mathrm{A},\mathrm{B}}=\left( \begin{matrix}
	0&		0&		0&		-1\\
	0&		0&		1&		0\\
	0&		1&		0&		0\\
	-1&		0&		0&		0\\
\end{matrix} \right) \left( \begin{matrix}
	1-C_{\mathrm{A},\mathrm{A}}-C_{\mathrm{B},\mathrm{B}}&		0&		0&		0\\
	0&		C_{\mathrm{A},\mathrm{A}}&		C_{\mathrm{B},\mathrm{A}}&		0\\
	0&		C_{\mathrm{A},\mathrm{B}}&		C_{\mathrm{B},\mathrm{B}}&		0\\
	0&		0&		0&		0\\
\end{matrix} \right) \left( \begin{matrix}
	0&		0&		0&		-1\\
	0&		0&		1&		0\\
	0&		1&		0&		0\\
	-1&		0&		0&		0\\
\end{matrix} \right) =\left( \begin{matrix}
	0&		0&		0&		0\\
	0&		C_{\mathrm{B},\mathrm{B}}&		C_{\mathrm{A},\mathrm{B}}&		0\\
	0&		C_{\mathrm{B},\mathrm{A}}&		C_{\mathrm{A},\mathrm{A}}&		0\\
	0&		0&		0&		1-C_{\mathrm{A},\mathrm{A}}-C_{\mathrm{B},\mathrm{B}}\\
\end{matrix} \right) 
\end{equation}
We then calculate the eigenstate of $R=\sqrt{\sqrt{\rho _{\mathrm{AB}}}\tilde{\rho}_{\mathrm{AB}}\sqrt{\rho _{\mathrm{AB}}}}$:
\begin{align}
    &\lambda _1=\sqrt{C_{\mathrm{A},\mathrm{A}}\!\:C_{\mathrm{B},\mathrm{B}}}+\sqrt{C_{\mathrm{A},\mathrm{B}}C_{\mathrm{B},\mathrm{A}}}
    \\
    &\lambda _2=\sqrt{C_{\mathrm{A},\mathrm{A}}\!\:C_{\mathrm{B},\mathrm{B}}}-\sqrt{C_{\mathrm{A},\mathrm{B}}C_{\mathrm{B},\mathrm{A}}}
    \\
    &\lambda _3=\lambda _4=0
\label{eqS:currentin}
\end{align}
Here we use the positive semi-definite properties of a density matrix: $C_{\mathrm{A},\mathrm{A}}\!\:C_{\mathrm{B},\mathrm{B}}\geqslant C_{\mathrm{A},\mathrm{B}}C_{\mathrm{B},\mathrm{A}}$. So we have the close form expression:
\begin{equation}
E_{\mathrm{C}}\left( \rho _{\mathrm{AB}} \right) =\lambda _1-\lambda _2=2\left| C_{\mathrm{A},\mathrm{B}} \right|
\end{equation}

Next, there is a relation between entanglement of formation and concurrence:
\begin{equation}
E_{\mathrm{F}}\left( \rho _{\mathrm{AB}} \right) =h\left( \frac{1+\sqrt{1-E_{\mathrm{C}}\left( \rho _{\mathrm{AB}} \right) ^2}}{2} \right) 
\end{equation}
So we have a closed-form expression of entanglement of formation: $E_{\mathrm{F}}\left( \rho _{\mathrm{AB}} \right) =h\left( \frac{1}{2}+\frac{1}{2}\sqrt{1-\left| 2C_{\mathrm{A},\mathrm{B}} \right|^2} \right)$, that can be deduced from coherence $C_{\mathrm{A},\mathrm{B}}$ directly. Note that in this paper we focus on entanglement of formation rather than concurrence mainly because we want to use entropy as an entanglement measure consistently.

\subsection{Relative entropy of entanglement}
In our model, bipartite entanglement has a closed-form expression to the coherence. Thus, a measure of entanglement can be equivalently expressed by the coherence measure, while a fully separable state is the incoherent state represented by a diagonal density matrix. Here we have the definition of relative entropy of entanglement:
\begin{equation}
    E_{\mathrm{R}}\left( \rho \right) =\min_{\rho _{\mathrm{sep}}} \left\{ -\mathrm{Tr}\left( \rho \ln \frac{\rho _{\mathrm{sep}}}{\rho} \right) \right\} ~,
\end{equation}
where $\rho _{\mathrm{sep}}$ run over all the diagonal matrix. Thus we assume that $\rho _{\mathrm{sep}}=\sum_{\alpha\in \left\{ \mathrm{g},\mathrm{p},1,...,N \right\}}{\Lambda _{\alpha}O_{\alpha}^{\dagger}|\mathrm{g}\rangle \langle \mathrm{g}|O_{\alpha}}$. Next, we have 
\begin{equation}
    -\mathrm{Tr}\left( \rho \ln \frac{\rho _{\mathrm{sep}}}{\rho} \right) =\mathrm{Tr}\left( \rho \ln \rho \right) -\mathrm{Tr}\left( \rho \ln \rho _{\mathrm{sep}} \right) =\sum_{\alpha}{\lambda _{\alpha}\ln \lambda _{\alpha}}-\sum_{\alpha}{C_{\alpha,\alpha}\ln \Lambda _{\alpha}}~,
\end{equation}
where $\lambda_{\alpha}$ are the eigenvalues of $\rho$. Consider the unit trace of a density matrix, $\sum_{\alpha}{\Lambda _{\alpha}}=\sum_{\alpha}{\lambda _{\alpha}}=1$, we can use the Lagrange multiplier 

\begin{align}
    &L\left( \left\{ \Lambda _{\alpha} \right\} ,l \right) =\sum_{\alpha}{\lambda _{\alpha}\ln \lambda _{\alpha}}-\sum_{\alpha}{C_{\alpha,\alpha}\ln \Lambda _{\alpha}}-l\left( \sum_{\alpha}{\Lambda _{\alpha}}-1 \right) ~,
    \\
    &\frac{\partial L\left( \left\{ \Lambda _{\alpha} \right\} ,l \right)}{\partial \Lambda _{\alpha}}=-\frac{C_{\alpha,\alpha}}{\Lambda _{\alpha}}-l\left( \sum_{\alpha}{\Lambda _{\alpha}}-1 \right) =0~,
    \\
    &\frac{\partial L\left( \left\{ \Lambda _{\alpha} \right\} ,l \right)}{\partial l}=\sum_{\alpha}{\Lambda _{\alpha}}-1=0~.
\end{align}
By solving these equations, we have the minimum of $-\mathrm{Tr}\left( \rho \ln \frac{\rho _{\mathrm{sep}}}{\rho} \right) $ reaches at $\Lambda _{\alpha}=C_{\alpha,\alpha}$, or
\begin{equation}
E_{\mathrm{R}}\left( \rho \right) =\mathrm{Tr}\left( \rho \ln \rho \right) -\sum_{\alpha\in \left\{ \mathrm{g},\mathrm{p},1,...,N \right\}}{C_{\alpha,\alpha}\ln C_{\alpha,\alpha}} ~.
\end{equation}

\newpage

\section{Supplementary Results}
In this section, unless otherwise stated, we use the following parameters to do the simulation: $N=40$, $\omega _{\mathrm{m}}=\omega _{\mathrm{c}}=2.11\mathrm{eV}$, $J=60\mathrm{meV}$, $\gamma _{\mathrm{decay}}=1.67\mathrm{THz}$, $\kappa=20\mathrm{THz}$, $\gamma _{\mathrm{dephasing}}=40\mathrm{THz}$, $\gamma _{\mathrm{in}}=\gamma _{\mathrm{out}}=1\mathrm{THz}$.

\subsection{Cavity enhanced transport}
Here we display the data $\left< I_{\mathrm{out}}\left( t \right) \right> $ as a function of $w$ and $G$ in a different way from the main text, to illustrate cavity enhanced transport better. We can also observe the disorder-enhanced transport occurs in the ultrastrong coupling regime.

\begin{figure}[htbp]
    \includegraphics[width=0.8\textwidth]{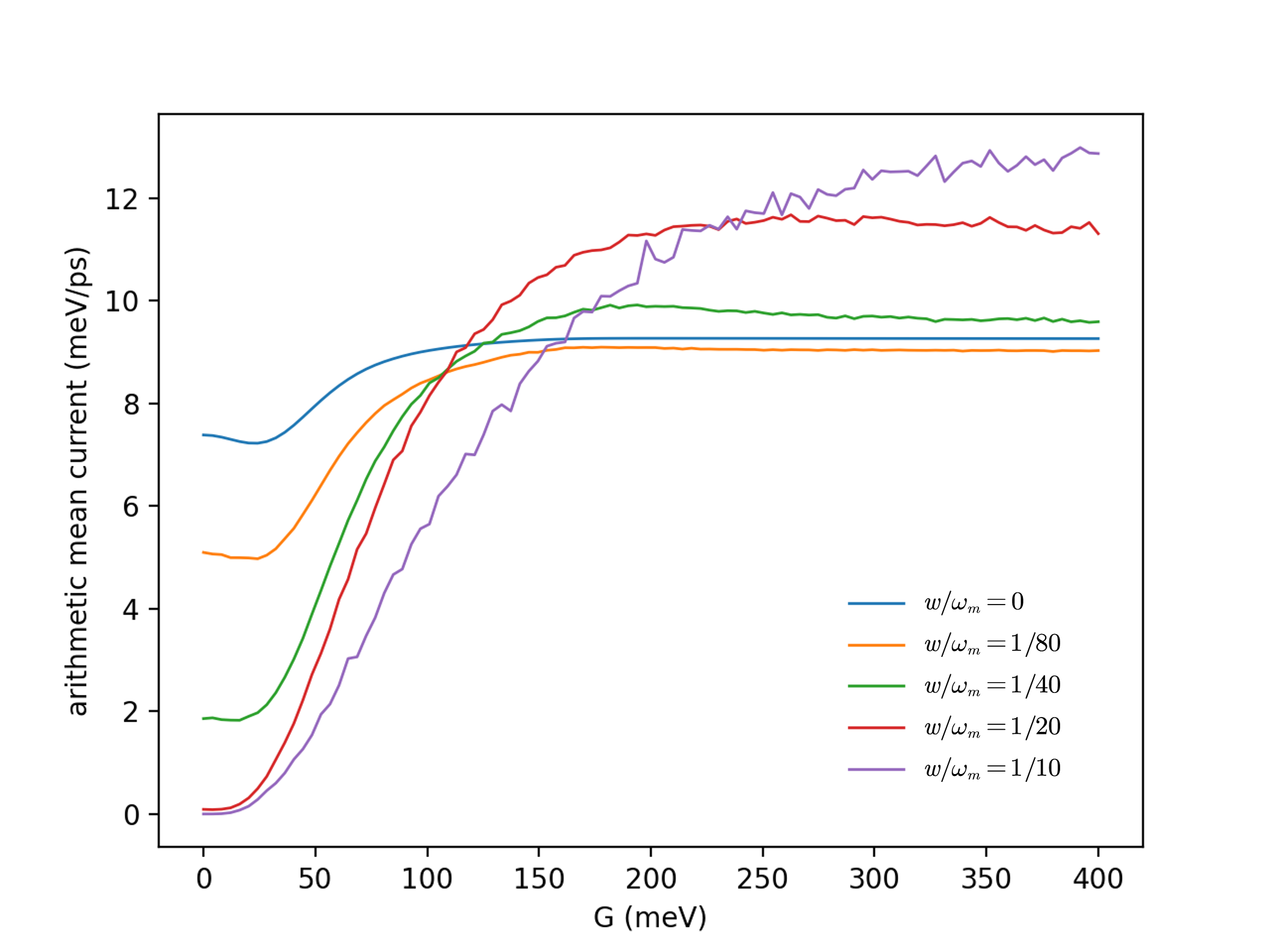}
    \caption{Cavity enhanced transport: $\left< I_{\mathrm{out}}\left( t \right) \right> $ as a function of $G$ for different disorder.}
    \label{figS:s7}
\end{figure}

\newpage

\subsection{Current statistics}
In this paper, we focus on the average current flowing into the drain $\left< I_{\mathrm{out}} \right>$, while the mean cannot always correctly reflect the distribution of each realization. Thus we need to check the Probability Density Function (PDF), based on sampling. We can see in both the disordered cavity case and disordered bare case, the PDFs show the near-gaussian distribution with a sharp peak. This implies that there is no long-tail effect. We can also check the difference between mean and median, in both cases they are close. This result supports our argument that the average $\left< I_{\mathrm{out}} \right>$ can reflect the statistics well. This means that for single realization, disorder-enhanced transport with the assistance of cavity QED is a high probability event.

\begin{figure}[htbp]
    \includegraphics[width=\textwidth]{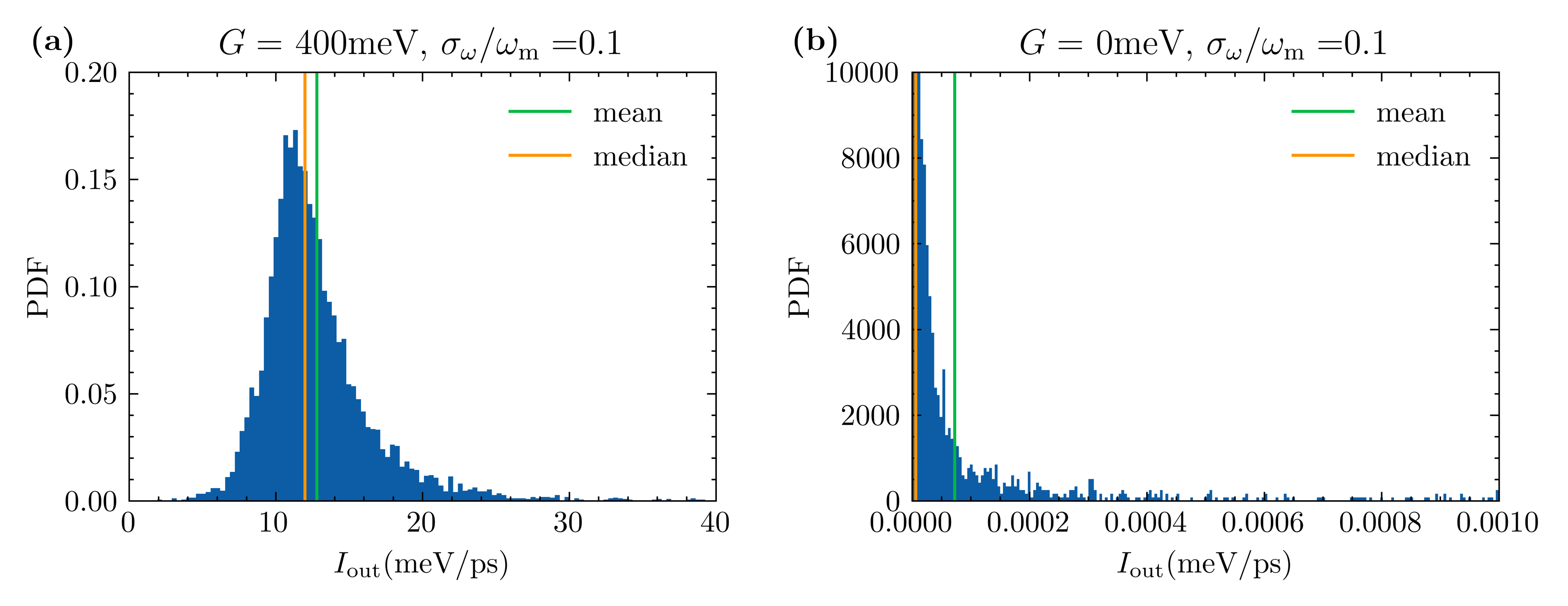}
    \caption{Statistics of (a) disordered and inside cavity ($w/\omega _{\mathrm{m}}=0.1, G=0\mathrm{meV}$) and (b) disordered and outside cavity ($w/\omega _{\mathrm{m}}=0.1, G=0\mathrm{meV}$).}
    \label{figS:s1}
\end{figure}

\newpage

\subsection{Matrix elements evloution}
We can explicitly expand the current $I_{\mathrm{out}}=\gamma _{\mathrm{out}}\mathrm{Tr}\left( H^{\mathrm{TC}}\mathcal{D} _{\sigma _{N}^{-}}\left[ \rho \right] \right)$ as
\begin{equation}
I_{\mathrm{out}}\left( G,\sigma _{\omega}/\omega _{\mathrm{m}} \right) =\gamma _{\mathrm{out}}\left( \omega _{\mathrm{m}}\left( \rho _S \right) _{N,N}-J\mathrm{Re}\left( \rho _S \right) _{N,N-1}+g\mathrm{Re}\left( \rho _S \right) _{N,p} \right)~,
\end{equation}
where the the density matrix elements including coherence between site $N$ and $N-1$, coherence between site $N$ and photon, and population of site $N$ contributes. We are also interested in the photon number to see the effect of the cavity. It can be seen that the disordered cavity system has the largest population terms.

\begin{figure}[htbp]
    \includegraphics[width=0.75\textwidth]{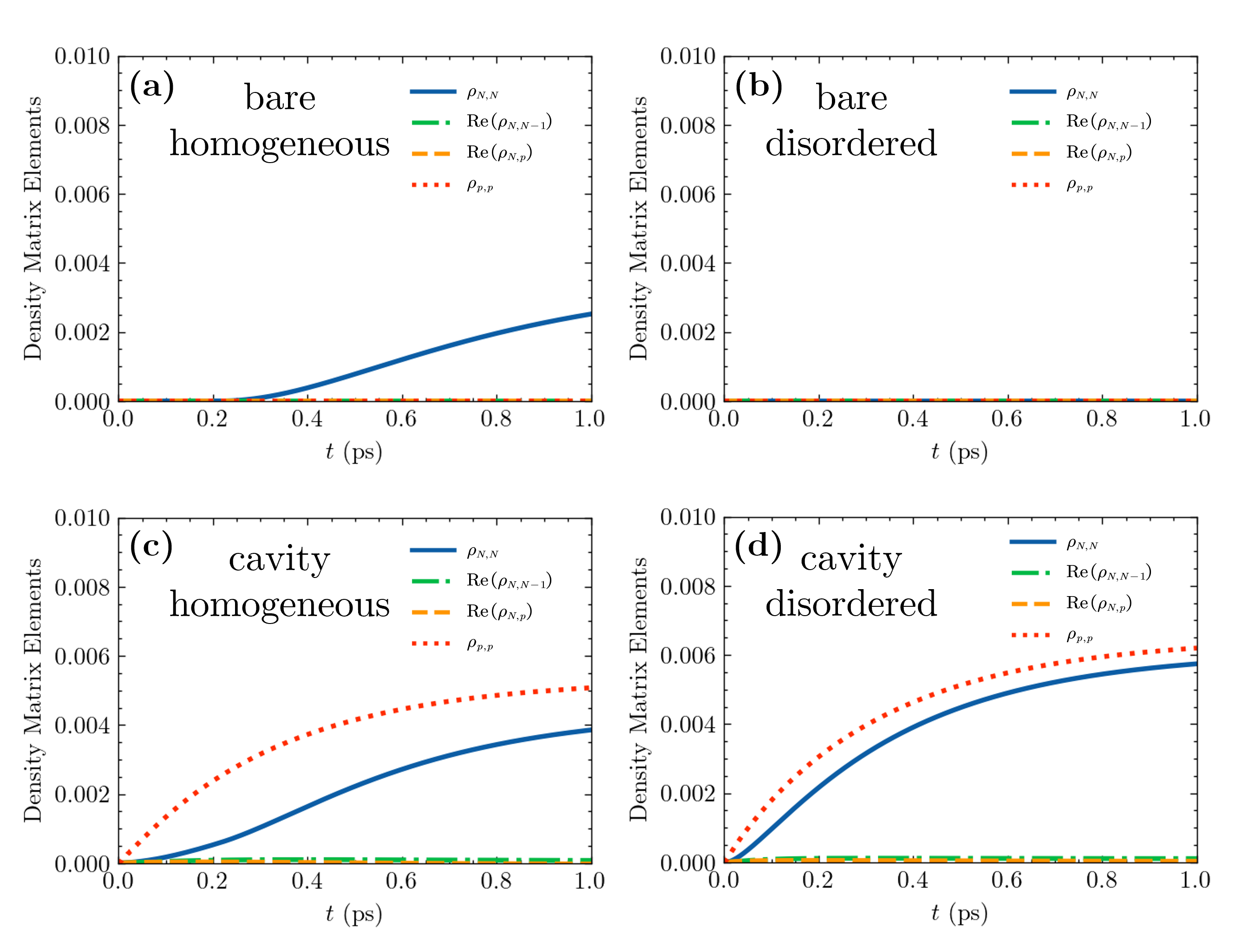}
    \caption{Ensemble average: Density matrix elements of (a) homogeneous and outside cavity ($w/\omega _{\mathrm{m}}=0, G=0\mathrm{meV}$), (b) disordered and outside cavity ($w/\omega _{\mathrm{m}}=0.1, G=0\mathrm{meV}$), (c) homogeneous and inside cavity ($w/\omega _{\mathrm{m}}=0, G=400\mathrm{meV}$), (d) disordered and inside cavity ($w/\omega _{\mathrm{m}}=0.1, G=0\mathrm{meV}$). }
    \label{figS:s2}
\end{figure}

\begin{figure}[htbp]
    \includegraphics[width=\textwidth]{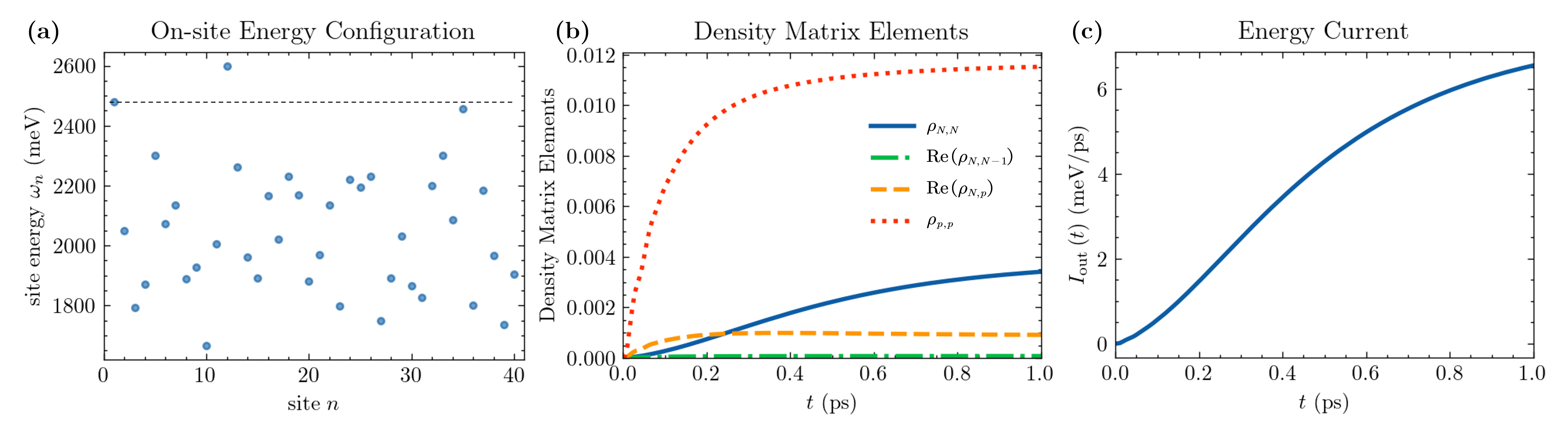}
    \caption{Single realization: (b)Density matrix elements and (c) current flowing into the drain of (a) specific energy configuration.}
    \label{figS:s3}
\end{figure}

\newpage

\subsection{Evolution of energy current and average exciton position}
We calculate the energy current $I_{\mathrm{out}}$ and average exciton position as functions of time. This shows that disorder-enhanced transport is not merely a late-time (steady state) effect, but also a phenomenon of the whole time regime. 

\begin{figure}[htbp]
    \includegraphics[width=0.9\textwidth]{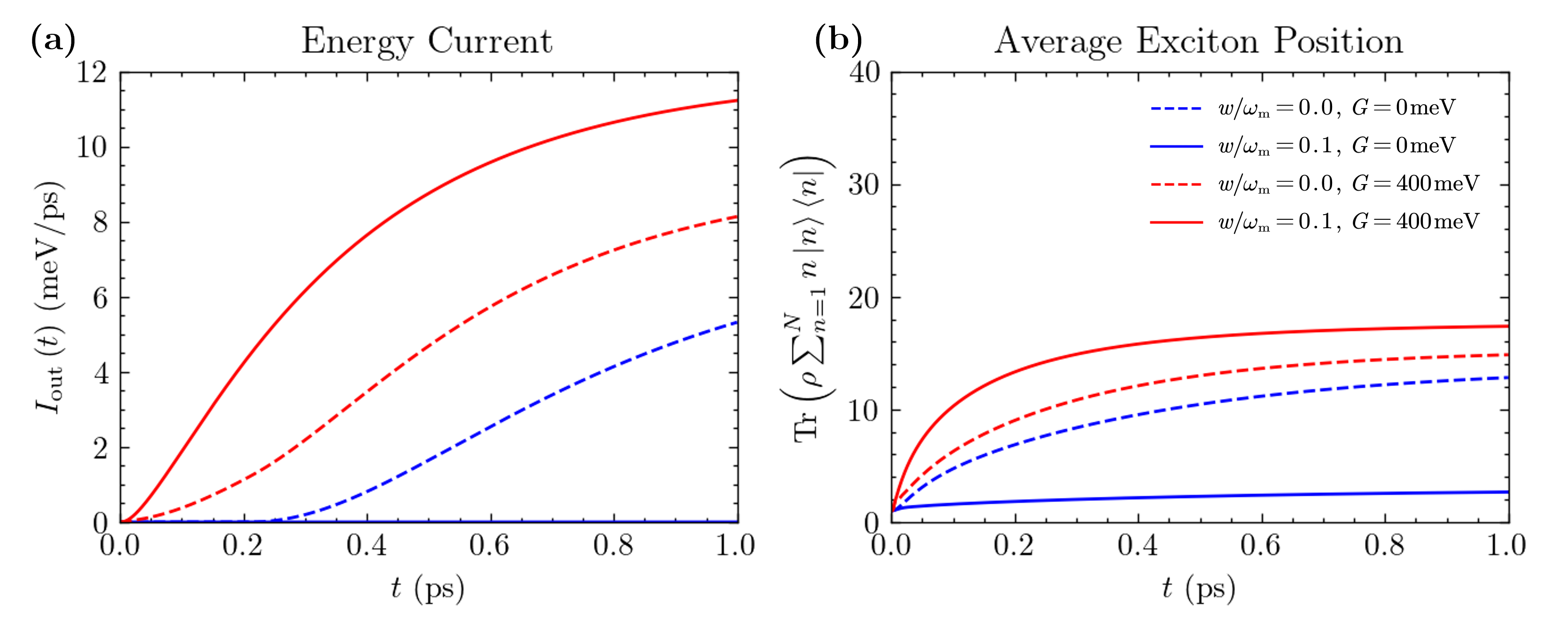}
    \caption{Evolution of energy current and average exciton position for the four cases, after ensemble average.}
    \label{figS:s4}
\end{figure}

\newpage

\subsection{Critical time}
Here we define the critical time as the population of site $N$ reaches a certain threshold $P_{\mathrm{cri}}$. There are two ways to study the critical time: (a) calculating the critical time for each ensemble and doing the average of each critical time. (b) calculating the critical time for the average population dynamics. It can be seem that in both cases, disorder-enhanced transport with the assistance of cavity QED in a certain range.

\begin{figure}[htbp]
    \includegraphics[width=0.9\textwidth]{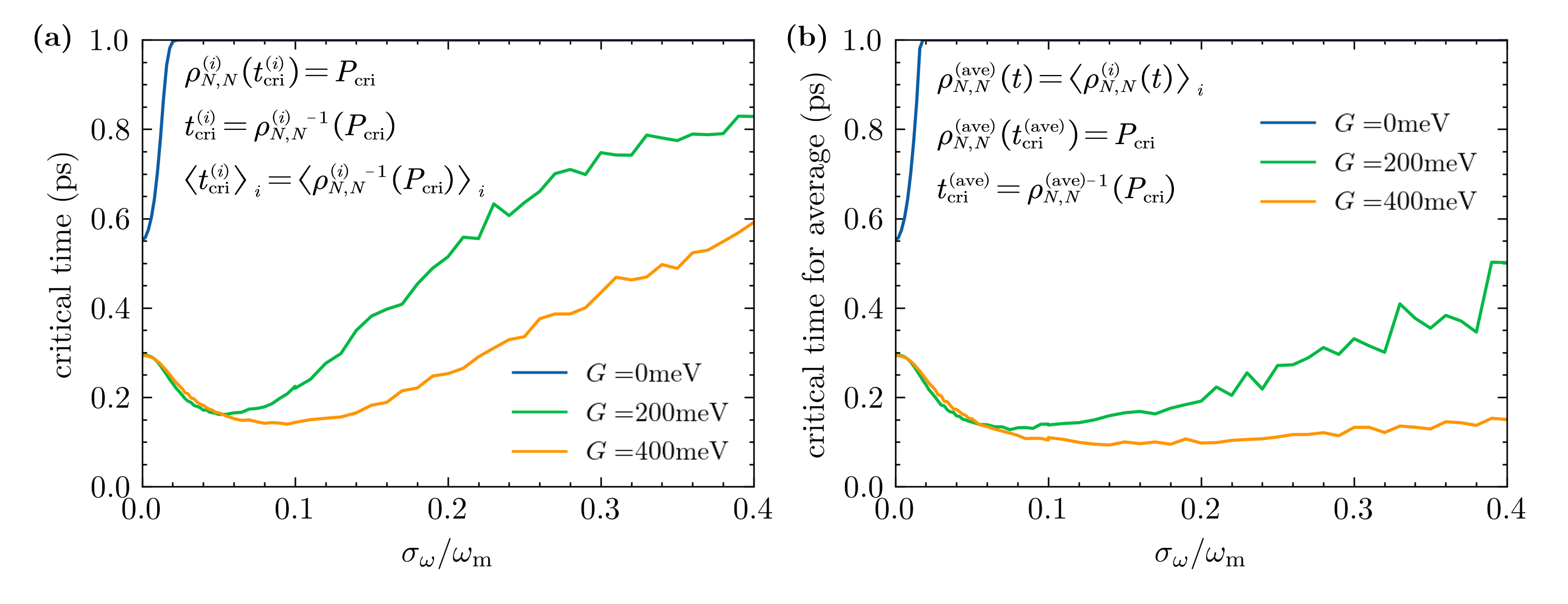}
    \caption{Critical time calculation with two methods.}
    \label{figS:s5}
\end{figure}

\newpage

\subsection{Classical Markov chain calculation}
To better understand the competition between the cavity-mediated jumping channel and site-to-site hopping channel, we build a classical Markov chain with a similar transport channel to study the dynamics. The dynamics can be calculated by the transfer matrix multiplication. Here we assume that the transfer rate obeys the Gaussian convolution:
\begin{align}
    &\textrm{cavity-mediated\ jumping}\quad k_{i\gets \mathrm{p}}=k_{\mathrm{p}\gets i}=K_{\mathrm{PE}}\left| g \right|^2e^{-\left( \frac{\omega _{\mathrm{c}}-\omega _i}{\sigma _{\mathrm{PE}}} \right) ^2}~,
    \\
    &\textrm{site-to-site\ hopping}\quad k_{i\gets i+1}=k_{i+1\gets i}=K_{\mathrm{EE}}\left| J \right|^2e^{-\left( \frac{\omega _{i+1}-\omega _i}{\sigma _{\mathrm{EE}}} \right) ^2}~,
\end{align}
where $K_{\mathrm{EE}}$, $K_{\mathrm{EP}}$, $\sigma_{\mathrm{EE}}$, $\sigma_{\mathrm{EP}}$ are four tunable parameters. $K_{\mathrm{EP}}$ represents the cavity-mediated jumping strength. $\sigma_{\mathrm{EP}}$ represents how Anderson localization suppresses cavity-mediated jumping. $K_{\mathrm{EE}}$ represents the site-to-site hopping strength. $\sigma_{\mathrm{EE}}$ represents how Anderson localization suppresses site-to-site hopping. In this setup, the steady state can be proved to be unique.

It can be seen that by tuning the parameters, we can barely see any disorder-enhanced transport with the optimal disorder $\sigma_{\omega}/\omega_{\mathrm{m}}=0.1$, implying the quantum effect plays a role in the disorder-enhanced transport. Note that for the arithmetic average, there are some meaningless negative results, implying that nonphysical variables emerge due to randomness.

\begin{figure}[htbp]
    \includegraphics[width=0.9\textwidth]{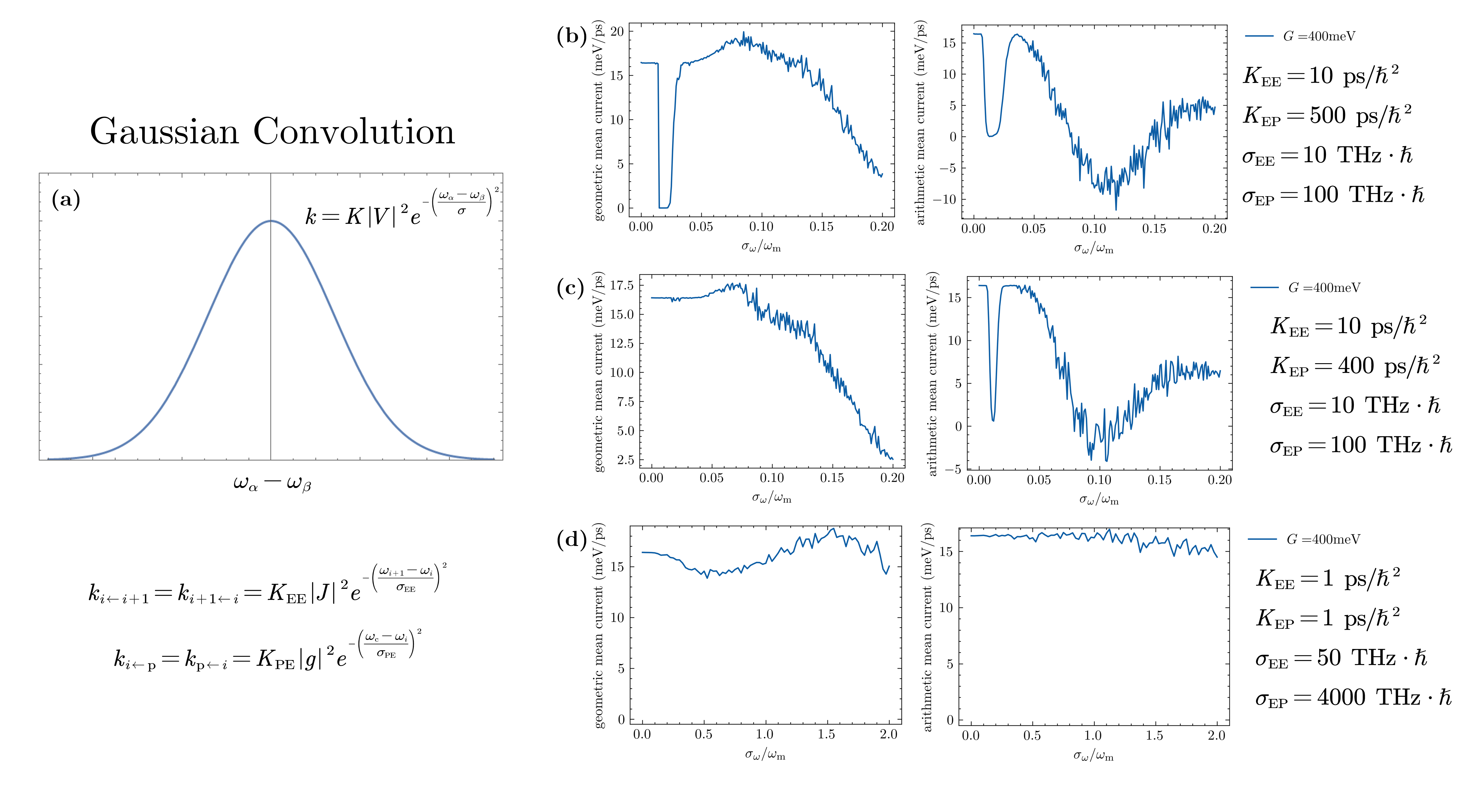}
    \caption{Classical Markov chain calculation for steady states. (a) an illustration of Gaussian convolution that rules the transfer rate. (b)(c)(d) are the calculations for different choices of $K_{\mathrm{EE}}$, $K_{\mathrm{EP}}$, $\sigma_{\mathrm{EE}}$, $\sigma_{\mathrm{EP}}$. In all (b)(c)(d), the the results of geometric average and arithmetic average are displayed.}
    \label{figS:s6}
\end{figure}

\end{document}